\documentclass[11pt,a4paper]{article}
\pdfoutput=1
\usepackage{hyperref}
\usepackage{amssymb}
\usepackage{graphicx,epsf,rotate,subfigure} 
\usepackage{times,cite,color,xspace}
\usepackage{multirow}
\usepackage{amsmath}
\usepackage{tikz}
\usepackage[normalem]{ulem}
\hypersetup{
pdfstartview=FitV,
colorlinks=true,
linkcolor=blue, 
citecolor=red, 
filecolor=black, 
urlcolor=blue}

\def\b{\beta}

\def\beq{\begin{equation}}
\def\eeq{\end{equation}}
\def\bea{\begin{eqnarray}}
\def\eea{\end{eqnarray}}

\def\bit{\begin{itemize}}
\def\eit{\end{itemize}}

\def\l{\left}
\def\r{\right}

\def\ra{\rightarrow}

\def\baa{\begin{array}}
\def\eaa{\end{array}}
\def\cQ{{\cal Q}}
\def \MET{\rm E{\!\!\!/}_T}

\def\dblone{\hbox{$1\hskip -1.2pt\vrule depth 0pt height 1.6ex width 0.7pt 
\vrule depth 0pt height 0.3pt width 0.12em$}}
\def\simgt{\mathrel{\lower2.5pt\vbox{\lineskip=0pt\baselineskip=0pt
\hbox{$>$}\hbox{$\sim$}}}}
\def\simlt{\mathrel{\lower2.5pt\vbox{\lineskip=0pt\baselineskip=0pt
\hbox{$<$}\hbox{$\sim$}}}}

\begin{document}
\begin{flushright}
{CERN-PH-TH-2015-10}
\end{flushright}
\begin{center}
{\Large \bf Same sign di-lepton candles of the composite gluons} \\
\vspace*{1cm} \renewcommand{\thefootnote}{\fnsymbol{footnote}} 
{ {\sf Aleksandr Azatov${}^{1}$}\footnote{email:aleksandr.azatov@cern.ch}, 
{\sf Debtosh Chowdhury${}^{2}$}\footnote{email:debtosh.chowdhury@roma1.infn.it},
{\sf Diptimoy Ghosh${}^{2,3}$}\footnote{email: diptimoy.ghosh@weizmann.ac.il} 
and {\sf Tirtha Sankar Ray${}^{4}$}\footnote{email:
tirthasankar.ray@gmail.com}}
\\
\vspace{10pt} {\small ${}^{1)}$ {\em Theory Division, Physics Department, 
CERN, Geneva, Switzerland} \\
${}^{2)}$ {\em INFN, Sezione di Roma, Piazzale Aldo Moro 2, I-00185 Rome, 
Italy } \\
${}^{3)}${\em Department of Particle Physics and Astrophysics,\\
Weizmann Institute of Science, Rehovot 7610001, Israel }
\\ ${}^{4)}$ {\em Department of Physics, Indian Institute of Technology 
Kharagpur, 721 302, India}}
\normalsize
\end{center}
%
\vspace{13mm}
\begin{abstract}
Composite Higgs models, where the Higgs boson is identified with the pseudo-Nambu-Goldstone-Boson 
(pNGB) of a strong sector, typically have light composite fermions (top partners) to account 
for a light Higgs. This type of models, generically also predicts the existence of heavy vector 
fields (composite gluons) which appear as an octet of QCD. These composite gluons 
become very broad resonances once phase-space allows them to decay into two composite 
fermions. This makes their traditional experimental searches, which are designed to look for narrow 
resonances, quite ineffective. In this paper, we as an alternative, propose to utilize the impact 
of composite gluons on the production of top partners to constrain their parameter space. 
We place constraints on the parameters of the composite resonances using the 8 TeV LHC data and 
also assess the reach of the 14 TeV LHC. We find that the high luminosity LHC will be able 
to probe composite gluon masses up to $\sim 6$ TeV, even in the broad resonance regime.
\end{abstract}
\setcounter{footnote}{0}
\renewcommand{\thefootnote}{\arabic{footnote}}
%
%
%
\clearpage
\tableofcontents
\section{Introduction}

The discovery of the Higgs boson at the Large Hadron Collider (LHC) \cite{:2012gk,:2012gu} has 
propelled us to the era of Higgs property measurements. Whether the discovered Higgs boson is an 
elementary or a composite object is an outstanding question, and would be at the cynosure of 
attention in the second run of the LHC which is about to start in a few months. In this context, 
models where the Higgs boson is a pNGB of a global symmetry {spontaneously broken} by a strongly coupled sector, 
represent well motivated scenarios of electroweak symmetry breaking containing a composite 
Higgs. \cite{Kaplan:1983fs, Georgi:1984af,Contino:2003ve} (see Ref.~\cite{Bellazzini:2014yua} 
for {a} recent review).

{In  the  models where the Standard Model (SM) fermion masses are generated by the partial 
compositeness mechanism \cite{Kaplan:1991dc}, the strong sector must contain 
fermionic colored resonances.
These, so called,  top partners,  are crucial to ensure the finiteness of the SM fermion contributions to the 
radiatively generated potential for the pNGB Higgs \cite{Agashe:2004rs, Contino:2003ve}. 
These resonances} are expected to be light in order to reproduce the observed mass of the SM Higgs 
boson without introducing additional tuning into the model 
\cite{Pomarol:2012qf,Matsedonskyi:2012ym,Marzocca:2012zn, DeCurtis:2011yx, Panico:2012uw}, and their 
direct search at the LHC \cite{Chatrchyan:2013uxa, Aad:2014efa} already constrains them to be 
heavier than $\gtrsim 800$ GeV.

Since the top partners are coloured, generically one expects the presence of coloured vector 
resonances as well. In this paper we focus on the indirect constraints on the composite 
vector fields (composite gluons) which are in the adjoint representation of $SU(3)_{\rm Color}$. 
They can be identified with the Kaluza-Klein (KK) excitation of the SM gluons in the five-dimensional 
realizations of the composite scenarios \cite{Contino:2003ve}. The two loop contribution of these 
composite gluons to the Higgs potential is known to soften the fine-tuning in these models 
\cite{Barnard:2013hka}. However, the low energy flavour violating observables, especially $\epsilon_K$ 
in the $K^0-\overline{K^0}$ system, were shown to strongly prefer the mass of the composite gluon to 
be $m_{\rho} \gtrsim 10-30$ {TeV  \cite{Csaki:2008zd,Casagrande:2008hr,
Blanke:2008zb,Agashe:2008uz},} thus making it impossible to produce them at the LHC. 
{Introduction of flavour symmetries \cite{Cacciapaglia:2007fw,Fitzpatrick:2007sa,Csaki:2008eh, Santiago:2008vq,
Delaunay:2010dw,Redi:2011zi}  can  make these vector resonances light while being compatible with the flavour observables.} 
In this work, however,  we will not rely on any additional symmetries in the flavour sector and 
assume that there are cancellations among different contributions, allowing the composite gluons 
to be light and hence, accessible at the LHC.

If the decay of the composite gluon to the top partners is kinematically allowed, typically large 
couplings of the strong sector imply that the composite gluon will have large width, comparable to 
its mass. In that case, the traditional approach to search for heavy gluons through resonance hunting 
may prove ineffective \cite{Carena:2007tn}. However, as we will elaborate in this work, these broad 
resonances can be cornered by several other (cut-and-count) searches being carried out at the LHC. 
In particular, the gluon partners contribute to the top partner pair production cross-section and this 
can be used to put useful constraints on them \cite{Carena:2007tn}\footnote{In principle this type of 
analysis can be used even for the narrow resonance searches however, if the resonance is within the  
kinematic reach bump-hunting may be a better search strategy.}. In this paper we adopt this approach 
and recast the 
studies carried out to search for top partners to constrain the composite gluon parameter space. 
{In particular, our study will focus on the indirect bounds on the parameter space of the composite gluons from the top partner searches with the same sign dilepton final state by the ATLAS \cite{ATLAS} and CMS \cite{CMS} collaborations\footnote{While we considered only the same-sign di-lepton channel in our analysis, there are 
also other channels (e.g., final state with { top quarks decaying hadronically}) which can be potentially important 
\cite{Backovic:2014uma,Azatov:2013hya}.}. We will also study the reach of the 14 TeV LHC. }
Recently this strategy was also used in the phenomenological study of Ref.~\cite{Chala:2014mma}, which 
however was focused on the parameter space with a narrower decay width of  the composite gluon. 
For some other related studies, we  refer the reader to Refs.~\cite{Agashe:2006hk,Lillie:2007yh,
Lillie:2007ve,Bini:2011zb,Kong:2011aa,Vignaroli:2015ama,Greco:2014aza}.


%
%
%

The rest of this paper is organized as follows. In Section \ref{model} we will present the 
Lagrangian of our simplified model and briefly discuss the branching ratios of the composite gluon 
to various top partners. In Section \ref{broad} we will discuss the subtleties involved in dealing 
with broad resonances. The details of our numerical simulation will be presented in section 
\ref{simulation}. We will present our main results in section \ref{results} and conclude thereafter.

\section{The model setup} \label{model} 

In this section we present the basic structure of our model. We assume that the global symmetry 
breaking pattern leading to the pNGB Higgs is given by the $SO(5)/SO(4)$ coset. This is the minimal 
coset that contains an unbroken custodial symmetry.  We will assume that the SM fermion masses are 
generated by the partial compositeness mechanism.  The simplified two-site construction 
\cite{Contino:2006nn} will be utilized to describe the phenomenology of the lightest composite 
resonances. In particular, we will be interested in the phenomenology of the fermionic top partners 
and the partner of the SM gluon - the composite gluon and we will ignore the rest of the composite 
resonances.  For concreteness we focus on the ${\bf M4_{5}}$ model presented in 
\cite{DeSimone:2012fs}, minimally extended by the inclusion of the composite gluon. In this setup 
the top {partners belong to} the $\bf{4}$ of $SO(4)$ appearing as a part of ${\bf 5}$ 
of $SO(5)$.\footnote{This is the minimal construction which has a custodial protection for the large 
modifications of the $Z\bar b_L b_L$ coupling \cite{Agashe:2006at}.}
The relevant Lagrangian is given by
\bea
\label{2site}&&{\cal L}^{\bf M4_5} \supset -M_{\cal Q} \bar {\cal Q}{\cal Q} + y f 
(\overline{\Psi}_L)^I U_{Ii} {\cal Q}_R^i + 
y c_2 f (\overline{\Psi}_L)^I U_{I5} t_R  ~ ,
\eea
where ${\cal Q}$ is the the composite multiplet in the   representation $\bf{4}$, the $\Psi_L$ 
contains the SM left-handed quark doublet and $U_{I}$ is the non-linear representation of the pNGB 
Higgs and $t_R$ is assumed to be a fully composite state. Generically the lightest state is the field with charge 
$5/3$. One of the interesting features of this particle is that it decays with $100\%$ branching 
ratio into the $tW$ final state which, after the further decay of the top quark, leads to the same 
sign di-lepton final state. This interesting feature was used recently in the experimental studies to 
put bound on the mass of the fermionic top partners, $M_{5/3}\gtrsim 800$ GeV \cite{CMS}. Note 
that this bound was obtained assuming only the QCD pair production of the charge $5/3$ field. Later 
it was realized that the electroweak single production of the charge $5/3$ field can also lead to 
the same final state, thus making the overall bound even stronger \cite{DeSimone:2012fs,Matsedonskyi:2014mna}.

In this paper we follow a very similar approach and study the constraints {from the additional mechanism} 
for pair production of the charge $5/3$ field namely, processes mediated by the composite gluons. 
Note that in the model {$\bf M4_5$} we have one state with charge $5/3$, one state with charge 
$-1/3$ and two top-like 
states with electric charge $2/3$. We will denote these states by $X_{5/3}$, $B_{-1/3}$, $T^1_{2/3}$ 
and $T^2_{2/3}$ respectively (see Appendix \ref{app:mchm} for the details of the model setup).

The interaction of the composite gluons can be read off from the {two-site} model of the 
Ref.\cite{Contino:2006nn} {and is given by,}
\bea
\label{gaugeint}
{\cal L}_{gauge} =&& g_{QCD}A_\mu \l(\bar \cQ
\gamma^\mu\cQ +\bar \Psi_L \gamma^\mu \Psi_L+\bar t_R \gamma^\mu t_R\r) \nonumber\\
&&+ \rho_\mu
\l[\sqrt{g_*^2-g_{QCD}^2}\l(\bar \cQ\gamma^\mu\cQ+ \bar t_R \gamma^\mu t_R\r)
-\frac{g_{QCD}^2}{\sqrt{g_*^2-g_{QCD}^2}}\bar \Psi_L \gamma^\mu \Psi_L\r]. \eea
Hence,  the interaction of the composite gluon $\rho$  in the limit  $g_*\gg g_{QCD}$ can be written as
\bea
  \approx && \rho_\mu \l[g_*\l(\bar \cQ\gamma^\mu\cQ+\bar t_R \gamma^\mu t_R\r) -\frac{g^2_{QCD}}{g_*}\bar
\Psi_L \gamma^\mu \Psi_L\r]. \eea 
Similarly   the couplings between the other SM fermions (which we assume to be elementary) and 
the composite gluon are equal to
\bea
-\frac{g_{QCD}^2}{\sqrt{g_*^2+g_{QCD}^2}}\approx -\frac{g_{QCD}^2}{g_*}.
\eea
One can see that
the coupling of the elementary fermions to the composite gluon is suppressed (compared to the coupling 
to the SM gluon) by the 
factor $\frac{g_{QCD}}{g_{*}}$   which can be calculated in the explicit warped five-dimensional 
models and is given by $\frac{g_{*}}{g_{QCD}}\sim \sqrt{\log \frac{M_{\rm Pl}}{\rm TeV}}\sim 6$ 
\cite{Gherghetta:2000qt}.
Note that  Eq.~\ref{gaugeint} is written in the two-site basis, before the diagonalization of the 
fermion mass matrix.
The elementary left-handed top quark mixes strongly with the composite sector due to the  $yf$ term  
in the Lagrangian, see Eq.~\ref{2site}, and it is
convenient to introduce the parameter (sine of the mixing angle between $t_L$ and composite fields 
in the absence of the electroweak symmetry breaking),
\bea
s_L\equiv\frac{f^2 y^2}{\sqrt{f^2 y^2 +M_{\cal Q}^2}}, 
\eea
to  measure of compositeness of the left-handed top.

Let us summarize  some basic properties of the composite gluons that are important for 
phenomenology \cite{Contino:2006nn}. {Throughout this paper we will assume that all the light quarks 
(except for the bottom) are elementary.} Thus, the dominant production of the composite 
gluon ($\rho$) will be by the process $q\bar q\ra \rho$ with the coupling constant 
$\frac{g_{QCD}^2}{\sqrt{g_*^2-g_{QCD}^2}}$. Once produced, $\rho$ will decay predominantly into 
composite states due to 
the large coupling constant $g_*$. {In this work we will assume that only the SM fermions of the 
third generation mix strongly with the composite sector}
\footnote{Generically the composite gluons can decay also to the partners of the light 
quarks thus reducing the $Br(\rho\ra \hbox{top partners})$, however, as shown in \cite{Azatov:2014lha}, 
even these fields (partners of light quarks) in the anarchic scenarios are coupled predominantly to the 
third generation SM fields, so the similar analysis will be relevant.}. 
{The channels contributing to the signal in the same sign di-lepton final state, with some typical values of the branching fractions are given by:}
\bea
\label{eq:process}
&pp\ra \rho\ra X_{5/3} \bar X_{5/3}(X_{5/3}\ra t W)\nonumber\\
&Br(\rho
\ra X_{5/3} \bar X_{5/3})\sim 0.2-0.25,~~ Br( X_{5/3}
\bar X_{5/3}\ra {\hbox{same sign leptons}})\sim 0.09
 \nonumber\\ 
&pp\ra
\rho\ra B_{-1/3} \bar B_{-1/3}(B_{-1/3}\ra t W)\nonumber\\
&Br(\rho \ra
B_{-1/3}\bar B_{-1/3})\sim 0.07-0.15,~~ Br( B_{-1/3}\bar B_{-1/3}\ra
        {\hbox{same sign leptons}})\sim0.09 
\nonumber\\ 
&pp \ra \rho\ra  T^{}_{2/3}\bar T^{}_{2/3} (T_{2/3}\ra t Z, T_{2/3}\ra t h), \nonumber\\
&Br(\rho\ra
T^1_{2/3} \bar T^1_{2/3})\sim 0.08-0.15, ~ Br(\rho\ra T^2_{2/3} \bar T^2_{2/3})\sim
0.2-0.25,\nonumber\\
&~ Br(T_{2/3} \bar T_{2/3}\ra {\hbox{same sign  leptons}})\sim 0.02.
\eea 
It can be seen that the same sign di-lepton (electrons and muons only) final state will get the dominant 
contributions from the $5/3$ and $-1/3$ fields. The contribution of the {top-like} fields is not negligible, 
however it is much smaller than the effect of the $5/3$ field and hence, {we ignore} them in our analysis.
{The other 
major branching ratios for the composite gluon  are:}
\bea
\label{branchings}
Br(\rho \ra T^2_{2/3} t)\sim
0.03-0.06,~~ Br(\rho \ra t t)\sim 0.11-0.2\nonumber\\
Br(\rho \ra B_{-1/3} b)\sim
0.01-0.05,~~Br(\rho \ra b b)\sim 0.01-0.07 . 
\eea

\section{Analysis strategy}
\label{broad}
As we mentioned before, the goal of our study is to find the constraints coming from composite gluon 
mediated contribution to the top partner pair production. However, owing to the large coupling the 
decay width of the composite gluon is often comparable to its mass 
%
%
in {the} large region of the parameter space we are interested in. This invalidates the approximation of 
a narrow Breit-Wigner resonance for computation of the cross section ({for earlier studies of the wide width effects of the composite gluon resonances, see Refs. \cite{Djouadi:2011aj,Barcelo:2011vk,Barcelo:2011wu}}). Indeed the partonic cross section 
is proportional to
\bea
\label{eq:true}
\sigma(\hat{s})\propto \frac{1}{(\hat{s}-M_\rho^2)^2+({\rm Im}[{\rm M}^2(\hat{s})])^2 } \, ,
\eea
{where $-i {\rm M}^2(\hat s)$ is the sum of all one-particle-irreducible insertions into the $\rho$ 
propagator.
In the  limit  $M_\rho \gg \Gamma_\rho$  the cross section is dominated by the on-shell $\rho$ 
exchange and Eq.~\ref{eq:true} reduces to the standard Breit-Wigner formula by substituting 
\bea
-{\rm Im}[ {\rm M}^2(\hat s )] \Rightarrow -{\rm Im}[{\rm M}^2(\hat s = M_\rho^2)] = M_\rho \Gamma_\rho  .
\eea}
In Appendix~\ref{sec:kinematics} we report the formula of ${\rm Im}[{\rm M}^2(\hat s)]$ and discuss 
the situation when the narrow width approximation is expected to fail.
%
%
Instead of performing the full simulation with the true propagator shown in Eq.~\ref{eq:true}, 
we have divided our analysis into two parts. At first, we numerically calculate the total cross 
section using the exact formula of the propagator for every point in the relevant parameter 
space of the model. In the next step, in order to calculate the cut acceptance efficiencies, we 
first perform Monte Carlo simulation (including parton shower and hadronization) using
Madgraph/Pythia (see the following section for more details) in the narrow width approximation. 
We then estimate the finite width effects on the cut acceptance efficiencies in the following way:
\bit
\item for every value of the composite fermion mass $M_X$ we calculate the cut acceptance 
efficiencies for various values of the mass and width of the composite gluon. We denote it by 
$\epsilon_{M_X}( M_\rho,\Gamma_\rho)$.
\item for every mass of the composite fermion $M_X$ we find the minimal efficiency by 
varying $(M_\rho,\Gamma_\rho)$,
\bea
\label{eq:eff}
\epsilon^{Min}_{M_X}=\hbox{Min}[\epsilon_{M_X}( M_\rho,\Gamma_\rho)]\, .
\eea
We use $\epsilon^{Min}_{M_X}$ as a conservative estimate of the cut acceptance efficiency for 
the process of pair production (via composite gluon exchange) of the heavy fermions with mass 
$M_X$.
\eit

Our procedure of estimating the efficiencies is well justified because of the fact that for a 
given value of the partonic center of mass energy $\sqrt{\hat s}$, the angular distribution of composite
fermion pair production is independent of whether the full propagator of Eq.~\ref{eq:true} 
or the narrow-width approximation is used. The difference is just an overall factor, because the 
modification in going from the former (true propagator of Eq.~\ref{eq:true}) to the latter 
(narrow Breit-Wigner resonance) is entirely a function of the kinematic variable $\hat s$. 
So the only modification will appear in the $\hat {s}$-distribution which can be estimated 
by studying the distributions for various values of $M_\rho$ and  $\Gamma_\rho$ 
(for a fixed $M_X$).
{In order to estimate the error of this approximation we will also provide a comparison of the approximate efficiencies  with the 
exact calculation for a few benchmark points. }

\section{Details of collider simulation}
\label{simulation}

In this section we briefly describe the steps followed to perform the simulation and the event 
selection criteria used in our analysis. We have implemented the model in FeynRules 2.0 
\cite{Alloul:2013bka} and created the corresponding UFO files for the Madgraph event generator 
\cite{Alwall:2014hca}. Madgraph 2.2.1 has been used to generate {the parton} level events.  
Subsequently the Madgraph-Pythia interface \cite{Sjostrand:2006za, Sjostrand:2007gs} was 
utilized to perform the {showering and hadronization} of the parton level events and implementing 
our event selection cuts. The {parton distribution} function CTEQ6L \cite{Pumplin:2002vw} has been 
used throughout our analysis. We have employed the Fastjet3 package 
\cite{Cacciari:2011ma,Cacciari:2008gp, Dokshitzer:1997in} for reconstruction of the jets and 
implementation of the jet substructure analysis used for the reconstructing the top quarks and 
$W$ bosons.

As our goal is to recast the CMS \cite{CMS} and ATLAS \cite{ATLAS} 
\footnote{While this work was close to its completion, a new analysis by the ATLAS collaboration 
appeared \cite{Aad:2015mba} which found a slightly stronger lower bound on the mass of charge 
$5/3$ top partner, $M_{5/3}\gtrsim 840$ GeV.}
searches for the charge-$5/3$ top-quark partners in the same sign di-lepton final state, we have 
tried to follow their event selection procedures as closely as possible. For completeness, we 
present the step-by-step details of our analysis in appendix~\ref{app1}.
%
%

We find that the cut acceptance efficiency varies in the range $0.019-0.028$ for both the ATLAS 
and CMS 8 TeV analyses (our efficiencies include the branching ratio of $W$ boson into 
leptons). For the 14 TeV analysis we find that the efficiency varies between $0.009$ and 
$0.013$. 
%
%
%
\section{Results} \label{results} 

In this section we will present the final results of our study.  We start by analyzing the current 
LHC constraints on the composite gluons. Both the ATLAS and CMS collaborations have reported the 
exclusion limits on the QCD pair production of the fermionic top partners. 
In order to constrain the heavy composite gluons, we recast their results in the following 
way: we consider that a point in the parameter space of the model is excluded if the number 
of events predicted by the model $N_{\rm model}$ is larger than the $95\%$ C.L. exclusion limit 
reported by the experimental collaborations. In our analysis we ignore the interference between 
the composite gluon mediated pair production and the SM gluon contribution. This is a good approximation 
since the cross section is dominated by the on-shell $\rho$ production and only the $q \bar q$ initial 
state contributes to the $\rho$ mediated processes\footnote{In our analysis we have ignored the contribution of the single production of the composite 
fermions studied in \cite{DeSimone:2012fs,Azatov:2013hya,Matsedonskyi:2014mna}}.

As we have argued in the previous section, in order to accurately calculate the total cross section 
due to the wide resonances one needs to know $\rm{Im}[M^2(\hat{s})]$ for all values of $\hat s$ 
and {\it not only} on the mass peak. This requires the full knowledge of the masses of the particles 
and their couplings in the range of interest of $\hat s$, which makes it impossible to obtain 
completely model independent constraints. In this paper, as mentioned before, we have decided to 
focus on the ${\bf M4_5}$ model, which is the simplest composite Higgs construction containing 
charge $5/3$ and $-1/3$ fields.
{The model given in Eq.~\ref{2site}-\ref{gaugeint} can be parametrized in terms of the five 
independent parameters, $M_\rho$, $M_{\cal Q}$, $s_L$ , $g_*$ and $f$. 
In our numerical simulations we set $f=764$ GeV, 
which corresponds to  $10\%$ fine-tuning of the electroweak symmetry breaking scale. 
The parameter $c_2$  is fixed by 
requiring that the correct top quark mass is reproduced (see Eq.\ref{eq:m45}). }

In our calculation we consider the same sign di-lepton state originating from the QCD and composite 
gluon mediated pair production of the charge $5/3$ and $-1/3$ fields and we ignore the sub-dominant 
contribution of the charge $2/3$ top partners.  

The QCD pair production cross section was calculated 
using HATHOR \cite{Aliev:2010zk} at NNLO. For the composite gluon mediated contribution, we however
used the Leading Order (LO) cross section. Note that, the higher order corrections to the QCD pair 
production lead to an increase in the pair production cross section (see, for example 
\cite{Cacciari:2008zb}) with the corresponding $K$-factors $\sim 1.5$. Assuming that a similar 
increase happens also for the composite gluon mediated contribution, our use of LO cross section 
gives an conservative estimate of the expected bounds.

\begin{figure}
\begin{center}
\includegraphics[scale=0.65]{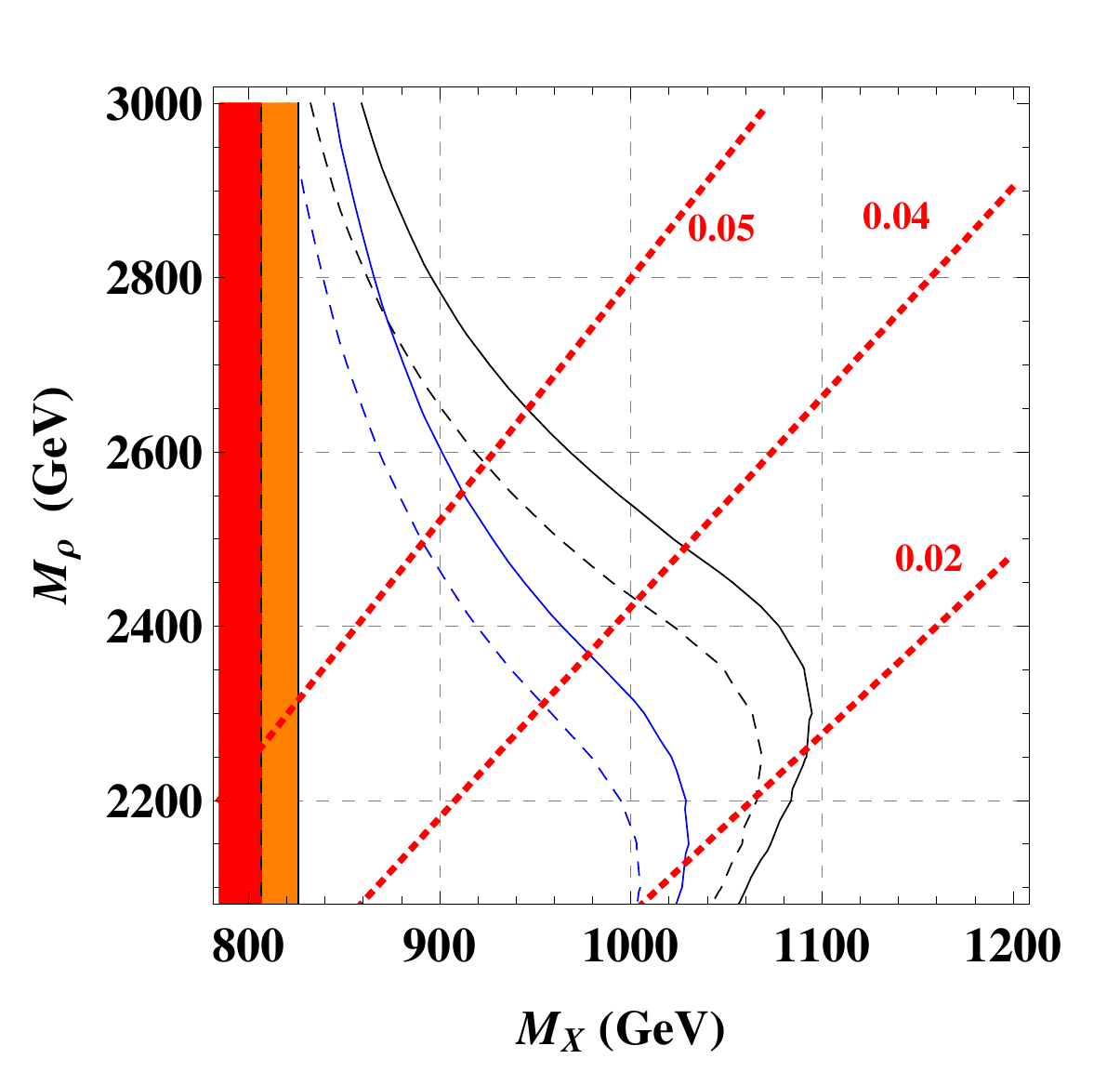}
\caption {{$95\%$ C.L. }exclusion contours in the $M_\rho - M_X$ plane for fixed value of $s_L = 0.5$. 
Solid lines represent the constraints obtained from recasting {the  CMS} study and the 
dashed lines correspond to the ones from the ATLAS study. The black (blue) lines correspond 
to the value $g_*=2.5(3)$. The red dotted lines indicate the ratio $\frac{\Gamma_\rho}{M_\rho}\times 
\l(\frac{1}{g_*^2-g_{QCD}^2}\r)$.
 The orange and red  vertical bands 
are the constraints from the CMS \cite{CMS} and ATLAS \cite{ATLAS} searches respectively assuming 
only QCD production.
\label{fig:mrhomX}}
\end{center}
\end{figure}

\begin{figure}
\begin{center}
\includegraphics[scale=0.6]{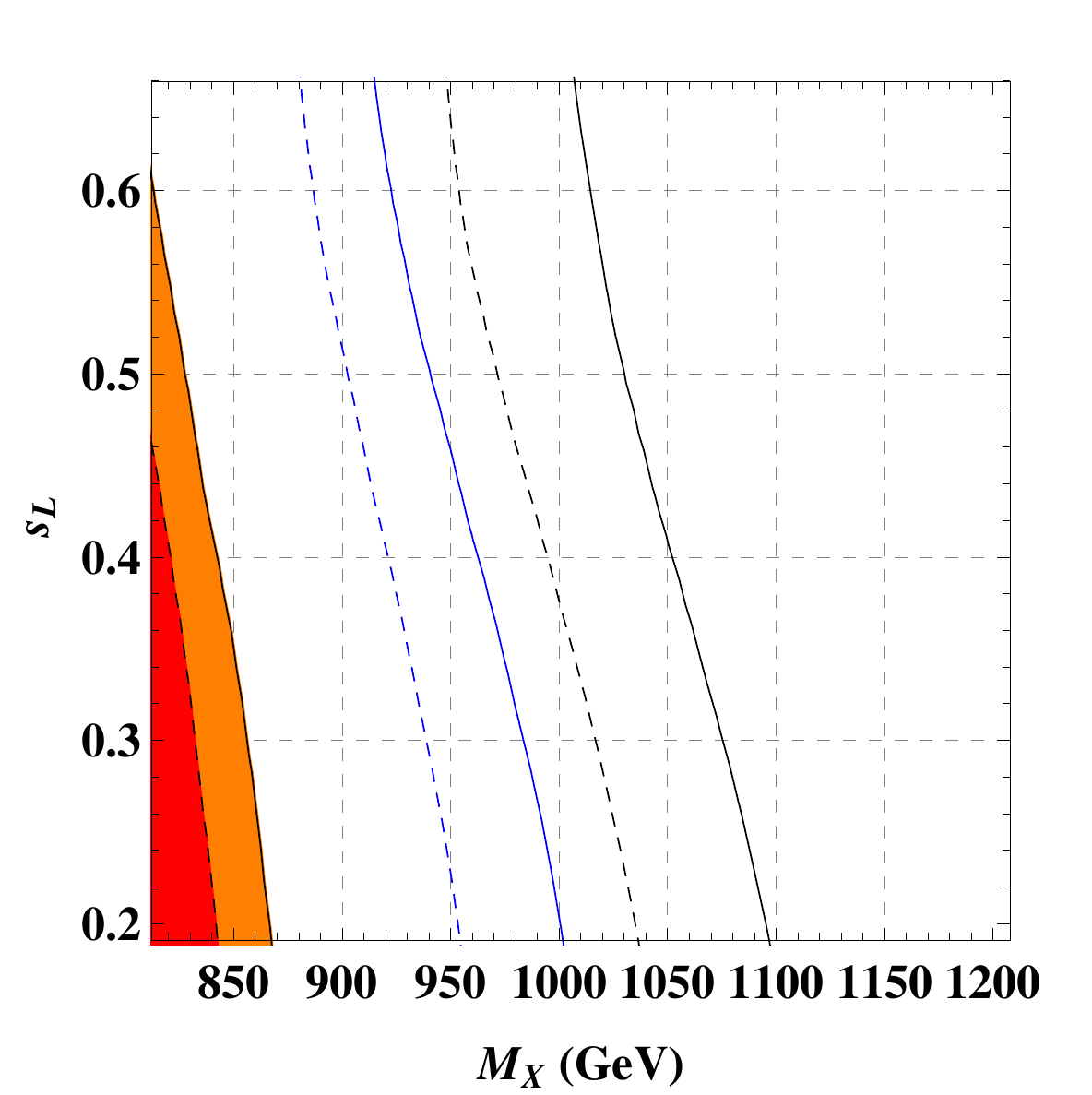}
\includegraphics[scale=0.66]{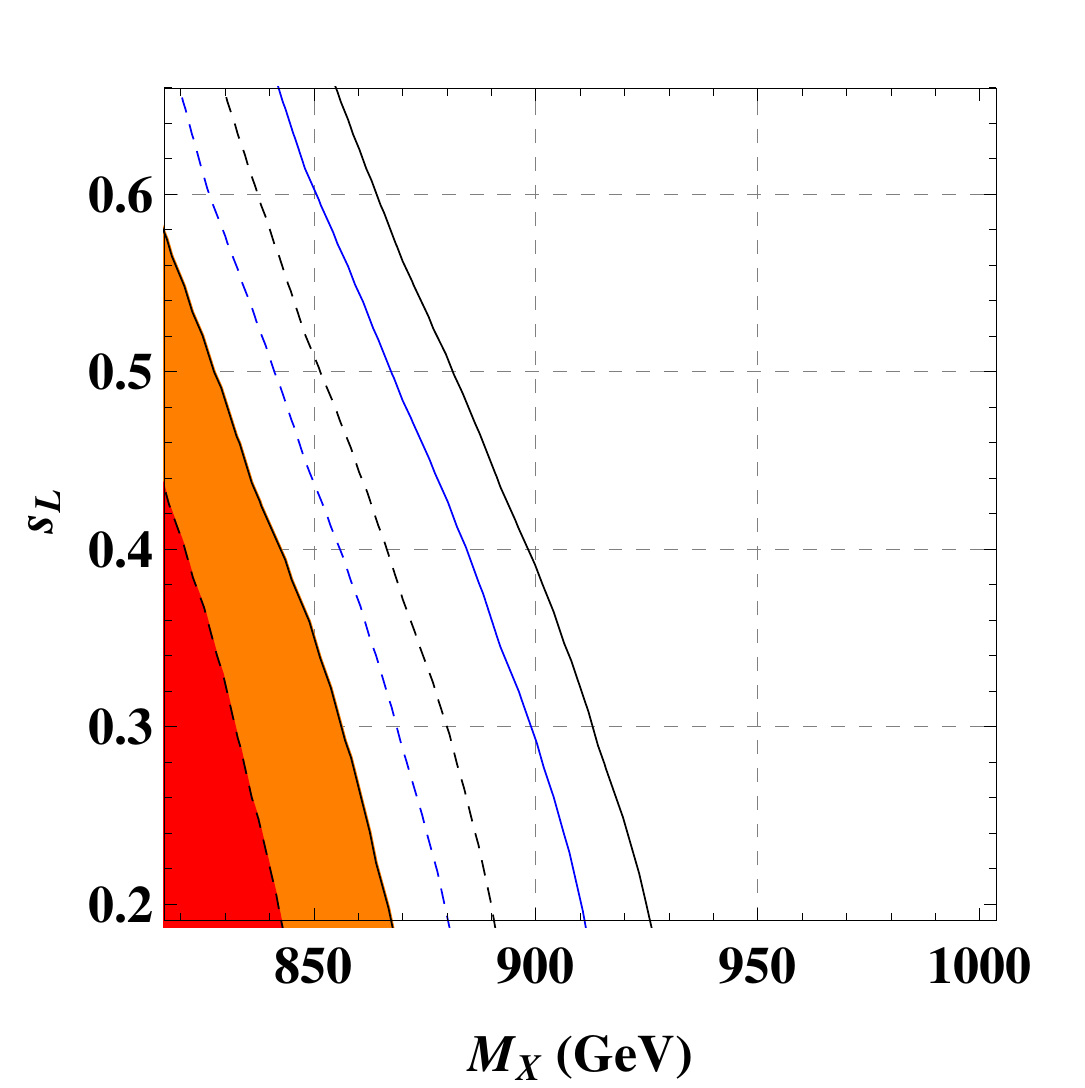}
\caption{ {95 $\%$ C.L.} exclusion contours in the $s_L - M_X$ plane for two fixed value of $M_\rho$, 
$M_\rho=2.5$ TeV (left panel) and $M_\rho=3$ TeV (right panel). 
Solid lines represent the constraints obtained from recasting the the CMS study and the 
dashed lines correspond to the ones from the ATLAS study. The black (blue) lines correspond 
to the value $g_*=2.5(3)$. The orange and red   bands 
are the constraints from the CMS \cite{CMS} and ATLAS \cite{ATLAS} searches respectively assuming 
only QCD production.
\label{fig:sLmX8}}
\end{center}
\end{figure}

\begin{figure}[h]
\begin{center}
\includegraphics[scale=0.65]{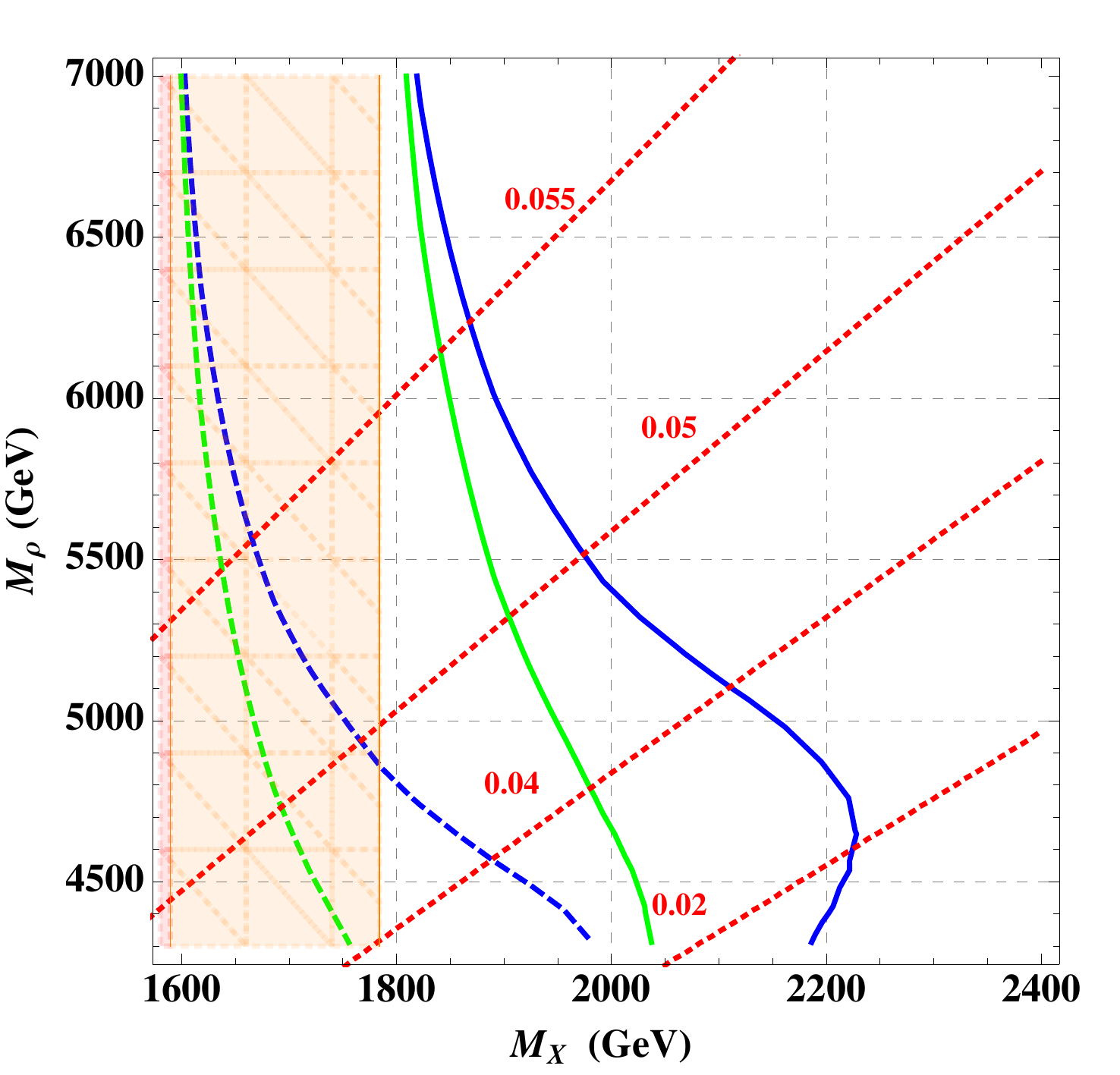}
\caption{{Prospects of the  $95\%$ C.L.} exclusion contours in the $M_\rho - M_X$ plane for fixed value of $s_L = 0.5$. 
The dashed (solid) lines represent LHC 14 exclusion reach for an integrated luminosity of 
300 fb$^{-1}$ (3 ab$^{-1}$). The blue (green) line corresponds to the value $g_*=3(4)$. 
The red dotted lines indicate the ratio $\frac{\Gamma_\rho}{M_\rho}\times 
\l(\frac{1}{g_*^2-g_{QCD}^2}\r)$.
 The red (orange) region is  
the exclusion prospects solely from QCD pair production at 300 fb$^{-1}$ (3 ab$^{-1}$).
\label{fig:mrhomX14}}
\end{center}
\end{figure}

\begin{figure}[h]
\begin{center}
\includegraphics[scale=0.61]{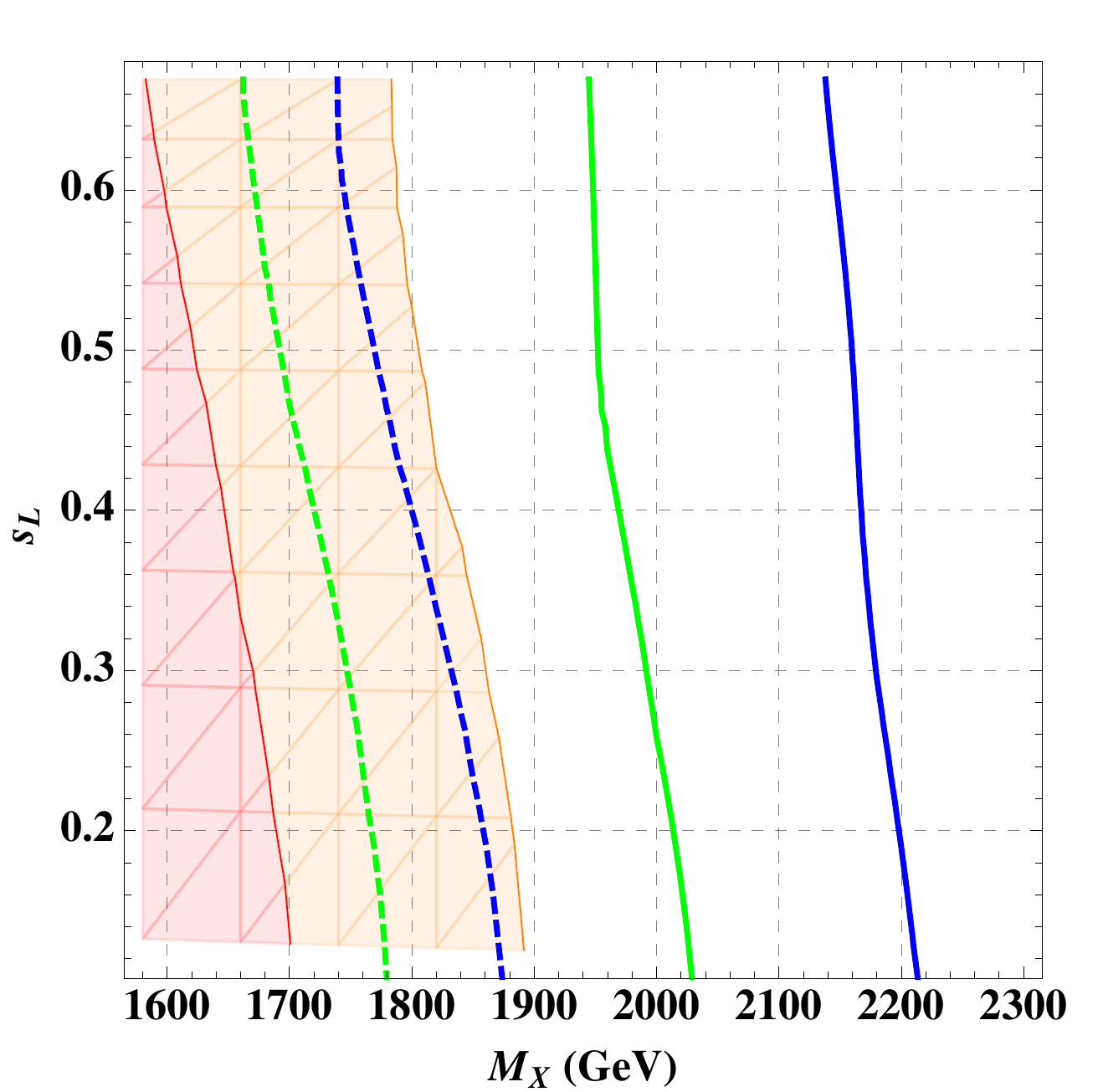}
\includegraphics[scale=0.65]{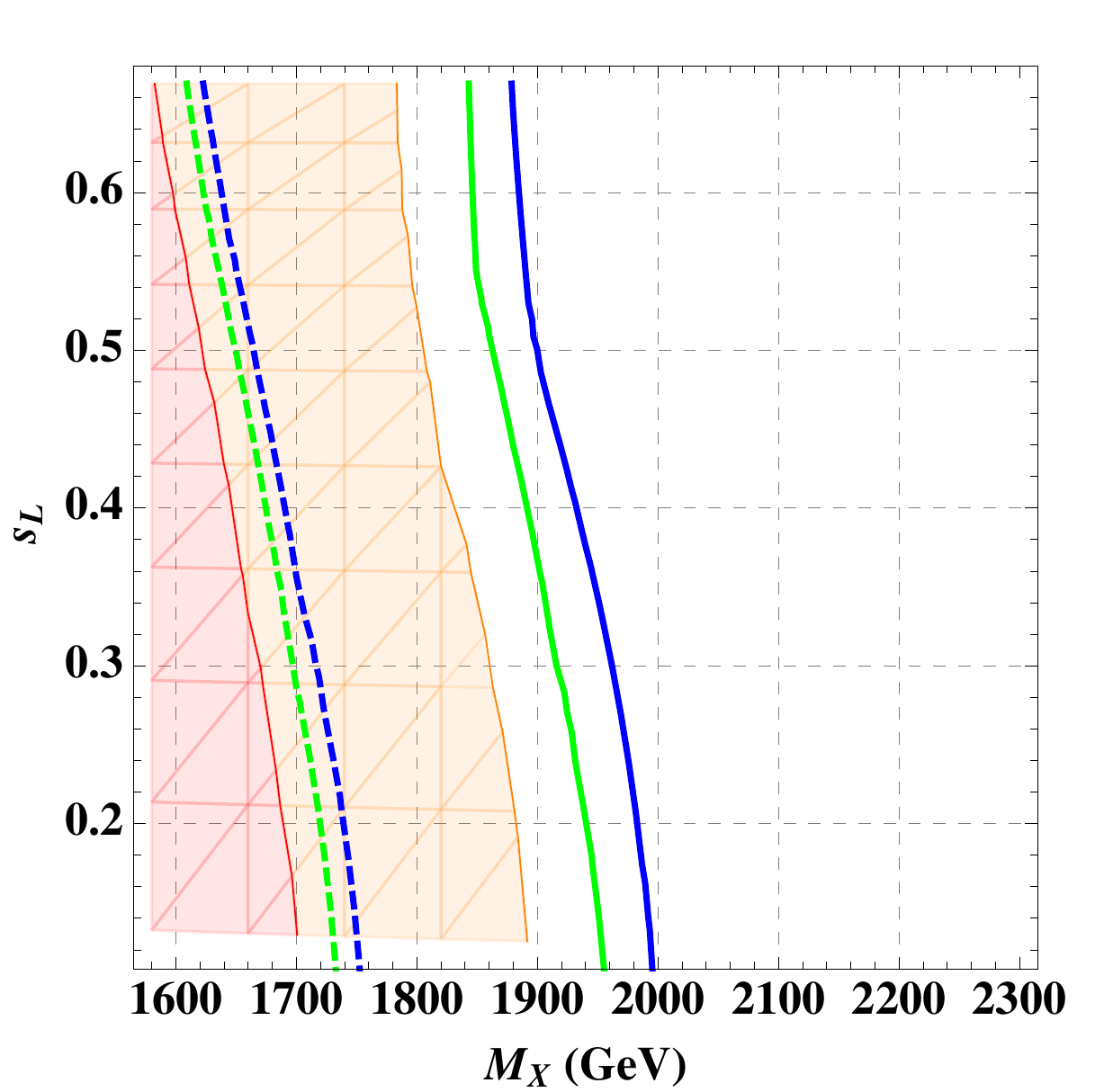}
\caption{{Prospects of the $95\%$ C.L.} exclusion contours in the $s_L - M_X$ plane for for two fixed value of $M_\rho$, 
$M_\rho=5$ TeV (left panel) and $M_\rho=6$ TeV (right panel).
The dashed (solid) lines represent LHC 14 exclusion reach for an integrated luminosity of 
300 fb$^{-1}$ (3 ab$^{-1}$). The blue (green) line corresponds to the value $g_*=3(4)$. 
The red (orange) region is the exclusion prospects solely from QCD pair production at 
300 fb$^{-1}$ (3 ab$^{-1}$).
\label{fig:sLmX14}}
\end{center}
\end{figure}

{ The exclusion plots  presented in the Fig.~\ref{fig:mrhomX}-\ref{fig:sLmX14} are obtained using the approximate efficiencies $\epsilon^{Min}_{M_X}$ defined in  Eq.~\ref{eq:eff}.
 However we crosschecked them against the true efficiencies for a few reference points using the modified version of the Madgraph/Pythia interface, where the full energy dependence of the composite gluon propagator was included. The results are presented in  the Table \ref{tab:compare}. One can notice that our method leads to a conservative estimate of the acceptance efficiencies and the difference between the true and approximate efficiencies is always within $25\%$, thus  justifying the use of the latter ones. }
\begin{table}[t]
\begin{center}
\begin{tabular}{|c|c|c|c|c|}
\hline
Reference points for 8 TeV LHC & CMS~$5/3$ & ATLAS~$5/3$ &CMS~$-1/3$ &ATLAS~$-1/3$  \\
\hline
$M_\rho=2.5 $ TeV,  $M_X=1$ TeV, $g_*=2.5$ &0.023(0.022) &0.023(0.023) &0.028 (0.024)& 0.025(0.023)\\
\hline
$M_\rho=2.5 $ TeV,  $M_X=0.9$ TeV, $g_*=3$ &0.025 (0.02)&0.023(0.022) &0.025 (0.22)&0.024 (0.023)\\
\hline
$M_\rho=2.2 $ TeV,  $M_X=0.9$ TeV, $g_*=3$ &0.024 (0.02)& 0.024(0.022)&  0.026(0.22)&0.024(0.023)\\
\hline
\end{tabular}
\begin{tabular}{|c|c|c|}
\hline
Reference points  for 14 TeV LHC & ~$5/3$ & $-1/3$  \\
\hline
$M_\rho=5.5 $ TeV,  $M_X=2$ TeV, $g_*=3$ &0.016 (0.013) &0.015 (0.012) \\
\hline
$M_\rho=5 $ TeV,  $M_X=2$ TeV, $g_*=4$ &  0.015 (0.013)&0.016(0.012)\\
\hline
$M_\rho=4.5 $ TeV,  $M_X=2$ TeV, $g_*=4$ & 0.015 (0.013)&0.016 (0.012)\\
\hline
\end{tabular}
\caption{The comparison between the true efficiencies and the $\epsilon_{M_{X}}^{Min} $(in parenthesis) defined in Eq.~\ref{eq:eff}. The mixing between left-handed top quark and composite fermions was  set   $s_L =0.5$ for all the reference points. \label{tab:compare}}
\end{center} 
\end{table}

{
Let us start by looking at the current bounds from the LHC searches.}
In Fig.~\ref{fig:mrhomX} we show the exclusion contours in the $M_\rho - M_X$ plane for a fixed 
value of left-handed top compositeness, $s_L = 0.5$. Similar exclusion contours in the 
$s_L - M_X$ plane are shown in Fig.~\ref{fig:sLmX8} for two fixed values of $M_\rho$, 
$M_\rho=2.5$ TeV and $M_\rho=3$ TeV. 
%
%
We find that the limits on the composite gluon mass relax substantially below the narrow 
width limit of 2.5 TeV \cite{Chatrchyan:2013lca,TheATLAScollaboration:2013kha} once {the decay 
channels into} the composite top partners becomes open. However, the current searches for the 
top partners still lead to interesting constraints on the composite gluons in the mass range of 
2-3 TeV, for the medium large composite gluon coupling $g_*\in[2, 3]$ and the width 
$\Gamma_\rho\sim (0.1-0.4) M_\rho $. For {the} smaller values of the coupling $g_*$ 
narrow resonance searches will become the most important tool in constraining the new colored 
resonances and for the larger $g_*$,  the composite gluon contribution becomes sub-dominant.

%
\textbf{LHC 14 TeV reach:} In order to estimate the discovery reach at the 14 TeV LHC we have 
adopted the analysis presented in Ref.~\cite{Avetisyan:2013rca}. We again present our results 
as $95\%$ C.L. exclusion contours in the $M_\rho - M_X$ plane (Fig.~\ref{fig:mrhomX14}) and 
$s_L -M_X$ plane (Fig.~\ref{fig:sLmX14}). It can be noticed that composite gluons up to the masses 
$\lesssim 6$ TeV and the fermions masses up to $\sim 2.1$ TeV can be probed for $g_*\sim 3-4$. {One can see from Fig.~\ref{fig:mrhomX14} that we can easily probe the composite gluons with the decay width  as large as 1 TeV,  the parameter space which is not covered by the narrow resonance searches. }

Even though the expected 14 TeV constraints are weaker than the current bounds from the flavour 
violating observables (e.g., $\epsilon_K$), one should notice that unlike flavour violating 
observables which scale as $\frac{1}{M_*^2}$ \cite{Agashe:2008uz} the collider constraint gets 
stronger for smaller values of $g_*$ (for fixed value of $M_*$), leading to complementary 
constraints for small/medium $g_*$.  
We would also like to comment that for a highly composite $t_L$ the left-handed bottom quark $b_L$ 
becomes composite as well and can give an important contribution to $pp\ra\rho$ process due to the 
bottom parton density function. However, we find that this 
effect can contribute at most as
\bea
\frac{\sigma(pp(b\bar b)\ra\rho)}{\sigma(pp(q\bar q)\ra\rho)}\lesssim 2\% \l(\frac{g_*}{4}\r)^4
\l(\frac{s_L}{\sqrt{2}}\r)^4\, , 
\eea
which is just a few percent for the values of the $g_*$ and $s_L$ we used in our study.

\section{Conclusion}

In this paper we have studied the collider phenomenology of the composite gluon within a composite 
Higgs model framework. Our study  {focused on the region of the parameter space} where the composite gluon is 
kinematically allowed to decay into two fermionic top partners. In this region, typically the width of 
the composite gluons is expected to be comparable to its mass thus rendering the traditional resonance 
searches less effective \cite{Chala:2014mma}. However, as pointed out in \cite{Carena:2007tn}, the 
contribution of the composite gluon to the pair production of two top partners can be still significant. 
In this context, we have studied the current constraints as well as high luminosity LHC prospects on the 
composite gluon using the additional contribution to the top partner pair production mediated by the 
heavy gluon. 
As the calculation of the composite gluon contribution to the top partner pair production cross section 
required knowledge of the full spectrum of the composite fields, {we have chosen  for simplicity to exclusively focus 
only on the 
model ${\bf M4_5}$}. In our analysis, we have calculated the total cross section treating 
carefully the finite width effects and we have performed a detailed collider simulations in order to find 
a conservative estimate of the cut acceptance efficiencies.

We found that while the current data probes the composite gluon in the mass range $2-3$ TeV, the high 
luminosity LHC will expectedly do much better and the exclusion limits can be extended to composite gluon 
masses of $\sim 6$ TeV approaching very close to the mass range motivated by the flavour physics 
constraints.

\section*{Acknowledgements} 
We would like to thank K.Agashe,  G.Panico and G.Perez  for comments on the manuscript and R. Contino 
for discussion.
We would like to acknowledge the hospitality of the Centro de Ciencias de Benasque Pedro Pascual 
where this project was envisaged. DC and DG acknowledge support from the European Research Council 
under the European Union's Seventh Framework Program (FP/2007-2013)/ERC Grant Agreement No. 279972. 
The work of TSR  is supported by INSPIRE faculty grant DST, Govt. of India and ISIRD grant IIT-Kharagpur, 
India.

\appendix

\section{Event selection criteria} 
\label{app1}

In this appendix we present the details of our cut-and-count analysis used for recasting the 
ATLAS \cite{ATLAS} and CMS \cite{CMS} 8 TeV results, and also the 14 TeV projection from 
\cite{Avetisyan:2013rca}.

\subsection{ATLAS : 8 TeV}

\begin{itemize}
\item  Selection-I  : An event is accepted only if it has exactly two leptons 
(electron or muon), both with the same electric charge.  
All leptons are selected with a transverse momentum cut $p_{T}^\ell \ge 24$ GeV 
and the pseudo-rapidity $|\eta| \le 2.4$. Moreover, the region of pseudo-rapidity 
$1.37 < |\eta| <1.52$ is excluded. Leptons are also required to satisfy the 
following isolation criteria, \\ 
(i) the distance between the lepton and any of the jets, $\Delta R (j \, \ell)$,  
must satisfy $\Delta R > 0.4$ (see below for the details of jet reconstruction) \\
(ii) the lepton should be far enough from all the other leptons, 
$\Delta R(\ell \, \ell) > 0.35$ \\
(iii)the ratio of the total hadronic transverse energy deposit within a cone of 
$\Delta R=0.35$ around the lepton to the lepton transverse energy is $\leq 
5\%$.

\item  Selection-II  : If the same sign leptons 
are of electron flavour {($e^+e^+$ or $e^-e^-$)}, their invariant mass 
($m_{ee}$) is required to satisfy  $m_{\ell\ell} > 15$ GeV 
and $|m_{\ell\ell}-m_Z| > 10$ GeV.

\item Selection-III : Jets are constructed using the anti-$\rm k_T$ 
\cite{Cacciari:2008gp} algorithm with the radius parameter $R$=0.4. Only those 
jets which satisfy $p_{T}^j \ge 25$ GeV and the pseudo-rapidity $|\eta| \le 2.5$ 
are selected. We demand the presence of at least 2 such jets in every event.  

\item Selection-IV : Every event is required to have at least one $b$-tagged jet.
A jet is identified as a $b$-jet if it is close ($\Delta R < 0.2$) to a 
$b$-quark. For the $b$-tagging efficiency ($\epsilon_b$) we use the prescription 
from reference \cite{Chatrchyan:2012paa} which gives  $\epsilon_b$ = 0.71 
for $90$ GeV $ < p_T < 170$ GeV and at higher (lower) $p_T$ it decreases linearly 
with a slope of -0.0004 (-0.0047) GeV$^{-1}$. Moreover, the probability of 
mis-tagging a $c$-jet (light jet) as a $b$-jet is taken to be 20\% (0.73\%) 
\cite{TheATLAScollaboration:2013xha}. 

\item Selection-V : We define the effective mass of an event to be 
$\rm m_{\rm eff} = \sum_{j} p_{T}^{j} + \sum_{\ell} p_{T}^{\ell}$ 
and demand that the event satisfies $\rm m_{\rm eff} > $ 650 GeV. Additionally, 
we also ask for a minimum missing transverse momentum $\MET > 40$ GeV 
in every event. 
\end{itemize}

\subsection{CMS : 8 TeV}

\begin{itemize}
\item  Selection-I  : An event is accepted only if it has exactly two leptons 
(electron or muon), both with the same electric charge. All leptons are 
selected with a transverse momentum cut $p_{T}^\ell \ge 30$ GeV and the 
pseudo-rapidity $|\eta| \le 2.4$. Leptons are also required to satisfy the 
following isolation criteria, \\ 
(i) the distance between the lepton and any of the reconstructed top quarks 
must satisfy $\Delta R > 0.8$ (see below for the details of top quark 
reconstruction) \\
(ii) the lepton should be far enough from all the other leptons, 
$\Delta R(\ell \, \ell) > 0.35$ \\
(iii) the ratio (the $\rm I_R$ threshold) of the total hadronic transverse 
energy deposit within a cone of 
$\Delta R=0.35$ around the lepton to the lepton transverse energy is $\leq 
17.5\%$.

\item  Selection-II  : If the same sign leptons are both electrons or both 
positrons, their invariant mass ($m_{ee}$) is required to satisfy 
$m_{ee} < 76$ GeV or $m_{ee} > 106$ GeV.

\item  Selection-III : We construct ``loose leptons'' with a lower $p_{T}^\ell$ 
cut of 15 GeV and relaxing the  $\rm I_R$ threshold to 50\%. Other selection 
criteria are kept identical to selection-I. We demand that all the same flavour 
opposite sign lepton pairs satisfy $m_{\ell \ell} < 76$ GeV or $m_{\ell \ell} 
> 106$ GeV.

\item  Selection-IV : The number of constituents ($N_c$) in each event should 
satisfy $N_c \geq 7$, where $N_c$ is defined as, 
\beq
N_c = N_j + N_\ell + 2N_W + 3N_t \, .
\eeq
Here $N_j$ is the number of jets which are constructed using anti-$\rm k_T$ 
algorithm with a distance parameter of 0.5 (AK5 jet).  
These jets are required to have $p_T > 30$ GeV and $|\eta| \le 2.4$. 
Moreover, they must be $\Delta R > 0.3$ away from the leptons in selection-I 
and $\Delta R > 0.8$ away from any other AK5 jet, reconstructed top quark and 
reconstructed $W$ boson. $N_\ell$ is the number of leptons counted from the 
same-sign di-leptons selected in selection-I. $N_W$ and $N_t$ refer to the total 
number of reconstructed top quarks and $W$ bosons respectively.

\item  Selection-V : $\rm m_{\rm eff} > $ 900 GeV where 
$\rm m_{\rm eff} = \sum_{j} p_{T}^{j} + \sum_{\ell} p_{T}^{\ell}$. 
All the jets in the definition of $m_{\rm eff}$ must be at least $\Delta R=0.3$ away 
from the selected leptons and $\Delta R=0.8$ away from any other jet.
\end{itemize}

In order to reconstruct the top quarks we used the Johns Hopkins top tagger
(JHTopTgger) \cite{Kaplan:2008ie} in our analysis. For the details of the 
steps followed in our simulation we refer the readers to section 3.2 of 
\cite{Biswas:2013hfa}. Here we briefly mention the differences compared to 
\cite{Biswas:2013hfa}. While constructing the fat-jets 
we used $R=0.8$. Unlike Ref.~\cite{Biswas:2013hfa}, we did not demand any 
limits on $\delta_p$, 
$\delta_r$ and cos$\theta_{h}$. However, the fat-jet is required to 
have $p_T > 400$ GeV and the pairwise invariant mass of the three highest 
$p_T$ subjets is required to be greater than 50 GeV. The invariant mass 
of the subjets is required to be roughly consistent with the top mass, within 
the range 100 GeV - 250 GeV.

In order to reconstruct the $W$ bosons we have used the algorithm proposed by 
Butterworth, Davison, Rubin, and Salam (BDRS) \cite{Butterworth:2008iy} to 
study the case of a light Higgs boson {($m_H \sim 125$ GeV)} produced in association with an 
electroweak gauge boson. For the step-by-step details of the algorithm, we refer 
our readers to Ref.~\cite{Byakti:2012qk}. The values of the parameters chosen in our 
analysis are exactly same to those used in Ref.~\cite{Byakti:2012qk} except the fact 
that we constructed the fat-jets with $R=0.8$ and asked for exactly 2 subjets 
in it. The fat-jets were also required to satisfy  $p_T > 200$ GeV. The 
two subjets should also have an invariant mass in the range 60 -100 GeV.

\subsection{ 14 TeV projections}

\begin{itemize}

\item  Selection-I   : At least two same-sign leptons with $p_{T}^\ell \ge 30$ 
GeV and the $|\eta| \le 2.4$. The leading $p_{T}$ lepton should also satisfy 
$p_{T}^\ell \ge 80$ GeV. While checking lepton isolation we only impose the 
criteria (ii) and (iii) mentioned in the previous section. 

\item  Selection-II  : Same as Selection-II in the previous section. 

\item  Selection-III : Same as Selection-III in the previous section.

\item  Selection-IV  : The number of constituents $N_c > 5$, where $N_c$ is 
defined in the same way as the previous section except that $N_\ell$ now refers 
to the number of leptons (with $p_{T} \ge 30$ GeV) excluding the two
leptons used for the same-sign di-lepton requirement.

\item  Selection-V : $\rm m_{\rm eff} > $ 1500 GeV where 
$\rm m_{\rm eff}$ is the scalar sum of the transverse momenta of all 
leptons and jets in the event with $p_T > 30$ GeV.  The missing transverse 
momenta $\MET$ and the sum $\rm m_{\rm eff} + \MET$ must also be more than 
100 GeV and 2000 GeV respectively. Moreover, the leading and the second 
leading jets in transverse momentum are required to satisfy $p_T >$ 150 GeV and 
50 GeV respectively. 
\end{itemize}

We have used the same prescription as detailed in the previous section to 
tag the top quarks and the $W$ bosons, the only difference being that the 
invariant mass of the subjets ($ \rm m_{inv}$) are now required to satisfy 
$ 140 < \rm m_{inv} < 230$ GeV and $60 < \rm m_{inv} < 120$ GeV for top tagged 
jets and $W$ tagged jets respectively.

\section{Kinematics}
\label{sec:kinematics}

In this section we will report some useful formulas for $\rho$ production and decay. We will assume 
that the part of the Lagrangian responsible for the production and decay of $\rho$ is given by,
\bea
\label{lag}
{\cal L} \supset g_{prod} \, \bar q \, t^a \gamma^\mu q \, \rho_\mu^a + 
g_{dec} \, \bar \chi_1 t^a \gamma_\mu(1+a_{12}\gamma_5) \chi_2 \rho_\mu^a \, \, .
\eea
The partonic cross-section of $\bar q q\ra \rho \ra {\bar \chi_1} \chi_2$ can be 
written as, 
\bea
\label{xsection}
&\hat\sigma(\hat s)=\dfrac{g_{prod}^2 \, g_{dec}^2}{54 \pi \hat s} 
\dfrac{\l(\hat{s}-(m_{\chi_1}-m_{\chi_2})^2\r)^{\frac{1}{2}}\l(\hat{s}-(m_{\chi_1}+
m_{\chi_2})^2\r)^{\frac{1}{2}}}{(\hat s- M_\rho^2)^2+
({\rm Im}[{\rm M}^2(\hat s )])^2}\times\nonumber\\
&\l\{(1+|a_{12}|^2)\l[ \hat s -\dfrac{\hat s (m_{\chi_1}^2+m_{\chi_2}^2)+(m_{\chi_1}^2-
m_{\chi_2}^2)^2}{2 \hat s}\r]+3m_{\chi_1}m_{\chi_2}(1-|a_{12}|^2)\r\} \, \, .
\eea
One can now compute the hadronic cross section in proton-proton collision 
which, following the standard notation, can be written as
\bea
\sigma_{had}=2\int_0^1 d \tau \hat \sigma ({S_{had}} \tau )\int_\tau^1 
\frac{dx}{x}\sum_{q}f_q(x) f_{\bar q}(\tau/x) \, \, ,
\eea
where the sum is over the parton distribution functions of all the light quarks inside the proton 
and the symmetry factor of two appears due to the  interchange of the  two partons in the initial state. 
In order to calculate the partonic cross section we need to know the $\rm{Im}[M^2(\hat s )]$ which, 
using the Cutkosky rules, can be written as,
\bea
-{\rm Im}[{\rm M}^2(\hat s)] = \dfrac{1}{2}\sum_f \int d\varPi_f |{\cal M}( \rho \to f)|^2 \, \, ,
\eea
where ${\cal M}( p \to f)$ is the matrix element of the process [$\rho \to $ final state $f$] 
assuming that $\rho$ has a mass $p^2=\hat s$ and $d\varPi_f$ is the corresponding phase space 
factor. For example, assuming that the resonance $\rho$ only decays to $\bar\chi_1 \chi_2$ fermions states the 
corresponding $\rm{Im }[M^2(\hat{s})]$ can be written as,
\bea
\rm{Im }[M^2(\hat{s})]_{\bar \chi_1 \chi_2}=&&-\frac{g_{dec}^2}{24\pi \hat{s}}
\theta({\sqrt{\hat{s}}}-m_{\chi_1}-m_{\chi_2})\l[(\hat{s}-(m_{\chi_1}+m_{\chi_2})^2)
(\hat{s}-(m_{\chi_1}-m_{\chi_2})^2)\r]^{1/2}\times \nonumber \\
&& \hspace*{-12mm}\l\{(1+|a_{12}|^2)\l[ \hat{s}-\frac{{\hat{s}} 
(m_{\chi_1}^2+m_{\chi_2}^2)+(m_{\chi_1}^2-m_{\chi_2}^2)^2}{2 \hat{s}}\r]+
3m_{\chi_1}m_{\chi_2}(1-|a_{12}|^2)\r\}\, .
\eea
%
%
\begin{table}[t]
\begin{center}
\begin{tabular}{|c|c|c|c|c|}
\hline
Ref. points & $M_\rho$, \, $M_{f}$, \, $g_{decay}$ &  decay width & $\sigma_{True}$ & $\sigma_{True}/\sigma_{NW}$ \\
\hline
A           & $M_\rho$=3 TeV, \, $m_\chi$=1.45 TeV, \, $g_{decay}$ = 5  & 374 GeV  &  0.033 pb  &  0.9\\
\hline
B           & $M_\rho$=3 TeV, \, $m_\chi$=1.2 TeV, \, $g_{decay}$ = 5 & 788 GeV & 0.061 pb & 1.25 \\
\hline
C           & $M_\rho$=3 TeV, \, $m_\chi$=1 TeV, \, $g_{decay}$ = 2 & 145 GeV & 0.079 pb & 1.01 \\
\hline
D           & $M_\rho$=3 TeV, \, $m_\chi$=1.495 TeV, \, $g_{decay}=$5  & 121 GeV & 0.029 pb & 0.42 \\
\hline
\end{tabular}
\caption{Comparison of the true cross section $\sigma_{True}$  with that obtained using narrow 
width approximation $\sigma_{NW}$ (where $-\rm{Im}[M^2(\hat{s})]$ was substituted by $\Gamma_\rho M_\rho$) 
for a few reference points. 
The coupling $g_{prod}$ has been set to unity. The hadronic center of mass energy was set to be 
equal to $\sqrt{S_{had}}= 8$ TeV.
\label{tab:ref}}
\end{center} 
\end{table}
\begin{figure}[h]
\begin{center}
\includegraphics[scale=0.7]{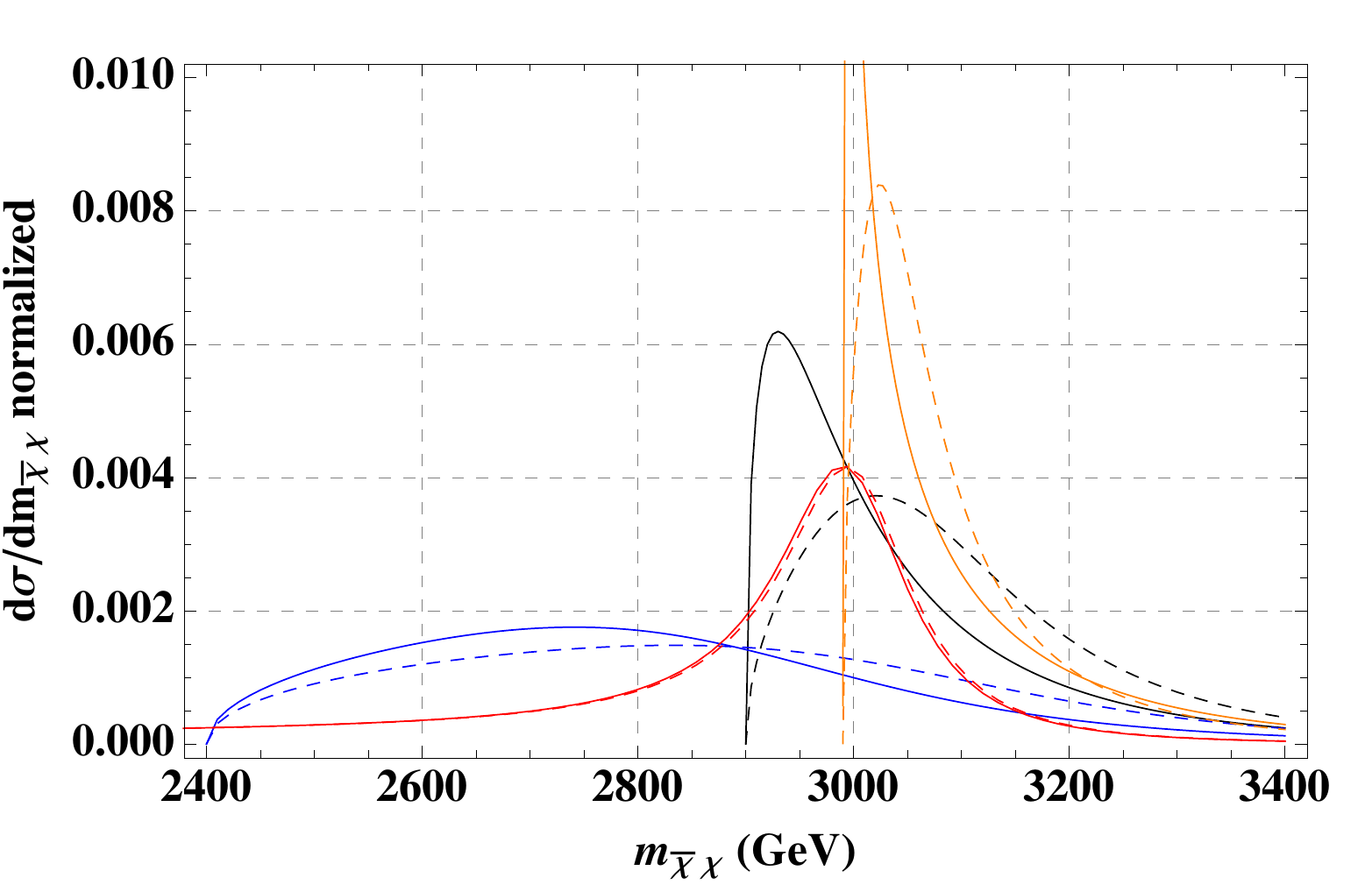}
\caption{ Normalized differential cross section as a functions of the invariant mass of the fermion 
pair, for $\sqrt{S_{had}}=8$ TeV collisions. 
{The black, blue, red and orange lines correspond to the reference points A, B, C and D respectively 
defined in table~\ref{tab:ref}. While the solid lines correspond to the true cross section, the dashed 
lines correspond to the fixed width approximation.} \label{fig:distr}} 
\end{center}
\end{figure}
In order to understand the effect of finite width of the composite gluon in a more quantitative way 
we  consider a simplified {model} where the $\rho$ 
{exclusively decays into the $ \bar \chi \chi$  pair of composite resonances  with the mass $m_\chi$.}
%
%
With this assumption, we compute the true production cross section as well as the one using 
fixed width approximation. The results are presented in the Table~\ref{tab:ref} and the Fig.~\ref{fig:distr}, where we show the differential cross section as a function of the invariant mass of the fermion 
pair. One can observe that the narrow width approximation is reproducing neither the shape nor 
the the integrated production cross section once the narrow resonance limit, 
$\Gamma_\rho\ll M_\rho$, is not satisfied. Another region where the narrow width approximation 
fails is near the threshold $2 m_\chi = M_\rho$. This can be understood by noticing that the total 
width vanishes above this threshold (i.e., $2 m_\chi > M_\rho$) unlike $\rm{Im}[M^2(\hat s)]$.

\section{Minimal composite Higgs model ${\bf M4_5}$}
\label{app:mchm}

In this section we briefly review the minimal composite Higgs model 
(we urge the interested readers to refer to the original literatures 
\cite{Agashe:2004rs,DeSimone:2012fs} for more details) which is based on 
the $SO(5)/SO(4)$ symmetry breaking pattern. In this paper we consider 
the composite gluon extension of the ${\bf M4_5}$ phenomenological model 
presented in \cite{DeSimone:2012fs}. The interaction between top quarks and 
composite fermions can be parametrized as
\bea
{\label{eq:m45}}
{\cal L}^{\bf M4_5}  \supset -M_{\cal Q} \bar {\cal Q}{\cal Q} + 
y f (\overline{\Psi}_L)^I U_{Ii} {\cal Q}_R^i + 
y c_2 f (\overline{\Psi}_L)^I U_{I5} t_R^5 
\eea
where ${\cal Q}$ is a multiplet (${\bf 4}$-plet of $SO(4)$) of the composite 
top partners, 
\bea
\cQ=\frac{1}{\sqrt{2}} \l(\baa{c}  i B_{1/3}-i X_{5/3}\\
B_{1/3}+X_{5/3}\\
iT^1_{2/3}+i T^2_{2/3}\\
-T^1_{2/3}+T^2_{2/3}
\eaa\r)
\eea
and $\Psi_L$ and $t_R$ stand for the SM fermions which are embedded into 
incomplete multiplets of $SO(5)$ namely, 
\bea
\Psi=\frac{1}{\sqrt{2}}\l(\baa{c}i b_L\\b_L\\it_L\\-t_L\\0\eaa\r) 
\, \, , \, \,
t_R^5 =\l(\baa{c} 0 \\ 0 \\ 0 \\ 0 \\ t_R \eaa\r) \, .
\eea

The $ 5 \times 5$ matrix $U$ containing the goldstone Higgs is given by 
{(in the unitary gauge)},  
\bea
U=\l(\baa{ccc} \dblone_3&&\\
& \cos \frac{h}{f}&\sin \frac{h}{f}\\
&-\sin \frac{h}{f}&\cos \frac{h}{f}
\eaa\r) \, \, .
\eea
The masses of the charge $5/3$ and $-1/3$ particles are given by,  
$M_{1/3}=\sqrt{M_{\cal Q}^2+y^2 f^2}$ and $M_{5/3}=M_{\cal Q}$ respectively.
{The masses of the  charge $2/3$  fermions  are given by the $3\times 3$
matrix
\bea
\l(\baa{ccc}
\frac{c_2 y f}{\sqrt{2}}\sin \frac{v}{f}& y f \cos^2 
\frac{v}{2f}& y f \sin^2 \frac{v}{2f}\\
0&-M_{\cal Q}&0\\
0&0&-M_{\cal Q}
\eaa \r),
\eea
where the lightest field is the SM top quark and the other two fermions have the masses  $M_{1,2/3}={\cal M_Q}$, 
$M_{2,2/3}=\sqrt{M_{\cal Q}^2+y^2 f^2}\, (1+{\cal O}(v^2/f^2))$}. 
The strength of the mixing between elementary and composite fields can 
be parametrized by the mixing angle 
\bea
s_L\equiv\sin \theta_L=\frac{y f}{\sqrt{y^2 f^2+M_{\cal Q}^2}} \, .
\eea

Extension of this setup by composite gluons was presented in 
\cite{Contino:2006nn} and the relevant part of the Lagrangian is given by 
\bea
{\cal L}_{QCD}=-\frac{1}{4}\rho_{\mu\nu}^2+\frac{F^2}{2}
(g_*\rho_\mu^*-g A_\mu^*)^2+g \bar {\cal E} \gamma^\mu {\cal E} A_\mu^* 
+g_*\bar {\cal C}\gamma^\mu {\cal C}\rho_\mu^*,
\eea
where $A^*$ and $\rho^*$ are the elementary and composite gluons respectively and ${\cal C, E}$ denote 
generic composite and elementary fermion fields. 
One can now find the mass eigenstates corresponding to the SM gluon $A_\mu$ and 
its partner $\rho_\mu$,
\bea
\rho_\mu=\frac{g_*\rho_\mu^*-g A_\mu^*}{\sqrt{g^2+g_*^2}},~~
A_\mu=\frac{g\rho_\mu^*+g_* A_\mu^*}{\sqrt{g^2+g_*^2}}.
\eea
The QCD interaction is given by the term 
\bea{
\frac{g g_*}{\sqrt{g^2+g_*^2}}A_\mu 
\l(\bar {\cal C}\gamma_\mu{\cal C} +\bar {\cal E}\gamma_\mu  {\cal E}\r)}
\eea
which gives, 
\bea{
g_{QCD} = \frac{g g_*}{\sqrt{g^2+g_*^2}} \approx g \, \, 
\hbox{in the limit} \, g \ll g_* \, \, .}
\eea
The couplings of the elementary and composite fermions to the 
heavy gluon can be written as
\bea
\label{eq:kkglue}
&& \rho_\mu
\l(\sqrt{g_*^2-g_{QCD}^2}\bar {\cal C}\gamma^\mu{\cal C}
-\frac{g_{QCD}^2}{\sqrt{g_*^2-g_{QCD}^2}}\bar  {\cal E} \gamma^\mu  {\cal E}\r) \nonumber\\
&& \approx \rho_\mu \l(g_*\bar {\cal C}\gamma_\mu{\cal C} -\frac{g^2_{QCD}}{g_*}
\bar  {\cal E} \gamma_\mu  {\cal E}\r).
\eea
Eq.~\ref{eq:kkglue} reveals that the composite gluon interacts with the composite fermion resonances 
with strength $\sim g_*$. Moreover, they will interact strongly also with the third generation SM 
fermions due to their strong mixing with the composite sector. The rest of the SM fermions couples 
to $\rho$ with a suppressed strength $g^2/g_*$. 

\end{document}